\documentclass[prd,nofootinbib,twocolumn]{revtex4}
\usepackage{graphics}
\usepackage{bm}
\usepackage{color}

\newcommand{\be}{\begin{equation}}
\newcommand{\ee}{\end{equation}}
\newcommand{\ba}{\begin{eqnarray}}
\newcommand{\ea}{\end{eqnarray}}

\newcommand{\nn}{\nonumber}
\newcommand{\Hi}{{\bm H}}
\newcommand{\D}{{\bm D}}
\newcommand{\T}{{\bm T}}
\newcommand{\g}{{\bm g}}
\newcommand{\h}{{\bm h}}
\newcommand{\A}{{\bm A}}
\newcommand{\vphi}{\vec{\phi}}
\newcommand{\vpsi}{\vec{\psi}}

\newcommand{\dslash}{\not{\hbox{\kern-2pt $\partial$}}}
\newcommand{\Dslash}{\not{\hbox{\kern-2pt D}}}
\newcommand{\kslash}{\not{\hbox{\kern-2pt $k$}}}
\newcommand{\nin}{\not{\hbox{\kern-5pt $\in$}}}
\newcommand{\Tr}{{\rm Tr}}

\begin{document}

\title{Domain walls and fermion scattering in Grand Unified models}

\author{D.~A.~Steer}
\email{steer@apc.univ-paris7.fr} \affiliation{APC\footnote{UMR
7164 (CNRS, Universit\'e Paris 7, CEA, Observatoire de Paris)}, 11
place Marcelin Berthelot, 75005 Paris, France \\ and \\
LPT, Universit\'e de Paris-Sud, B\^at.~210, 91405 Orsay Cedex,
France.}

\author{T.~Vachaspati}
\email{tanmay@monopole.phys.cwru.edu} \affiliation{CERCA,
Department of Physics, Case Western Reserve University, 10900
Euclid Avenue, Cleveland, OH 44106-7079, USA.}

\date{\today}

\begin{abstract}
\noindent Motivated by Grand Unification, we study the properties
of domain walls formed in a model with $SU(5)\times Z_2$ symmetry
which is spontaneously broken to $SU(3) \times SU(2) \times
U(1)/Z_6$, and subsequently to $SU(3) \times U(1)/Z_3$. Even after
the first stage of symmetry breaking, the $SU(3)$ symmetry is
broken to $SU(2)\times U(1)/Z_2$ on the domain wall. In a certain
range of parameters, flux tubes carrying color- and hyper-charge
live on the domain wall and appear as ``boojums'' when viewed from
one side of the domain wall. Magnetic monopoles are also formed in
the symmetry breaking and those carrying color and hyper-charge
can be repelled from the wall due to the Meissner effect, or else
their magnetic flux can penetrate the domain wall in quantized
units.  After the second stage of symmetry breaking, fermions can
transmute when they scatter with the domain wall, providing a
simpler version of fermion-monopole scattering: for example,
neutrinos can scatter into d-quarks, leaving behind electric
charge and color which is carried by gauge field excitations
living on the domain wall.

\end{abstract}

\maketitle

\renewcommand{\thefootnote}{\fnsymbol{footnote}}


\section{Introduction}

The interactions of fermions with non-trivial scalar field
profiles such as bubble walls, domain walls and other topological
defects have been the subject of numerous studies (see
\cite{Trodden:1998ym,ViSh} for reviews).
Depending on the model, a Yukawa-type coupling may lead to
localized zero modes (and hence to the existence of current
carrying cosmic strings for example), and it may also give
non-trivial fermion scattering. Indeed, the calculation of the
different reflection coefficients for particles and antiparticles
scattering of electroweak bubble walls was central to the picture
of electroweak baryogenesis put forward in the '80s
\cite{Trodden:1998ym}. Though this picture is no longer valid, in
\cite{Brand} a similar mechanism was revived using embedded
electroweak domain walls stabilized by thermal effects.

In this paper we study the interactions of bosons and fermions with
domain walls in an $SU(5)$ model motivated by grand unification.
Along the way we come across several novel phenomena which have not
seen much discussion in the literature. We discover that magnetic
monopoles carrying QCD color and hypercharge may interact
non-trivially with the wall since $SU(3)$ color and $U(1)$
hypercharge are broken in the wall and vortices exist there. We
also see that the scattering of fermions off the wall can lead to
baryon and lepton number violation. This is not so surprising
given that these numbers are not conserved in grand unified
theories. However, the scattering can also deposit electric charge
and $SU(3)$ charge on the domain wall when, for example, a
neutrino is transmitted through the wall as a d-quark. Gauge field
condensates carry this charge on the wall.

The domain walls are formed in the first stage of the symmetry
breaking scheme
\begin{eqnarray}
SU(5)\times Z_2 &\stackrel{\Hi}{\longrightarrow}&
\frac{SU(3) \times SU(2) \times U(1)}{Z_6} \nonumber \\
&\stackrel{\vphi}{\longrightarrow}& \frac{ SU(3) \times U(1)}{Z_3}
\label{basic.symbreak}
\end{eqnarray}
where $\Hi$ is a 24 component $SU(5)$ adjoint Higgs while $\vphi$
is a 5-component complex Higgs transforming in the fundamental
representation of $SU(5)$. The vacuum manifold for the first
symmetry breaking is disconnected due to the $Z_2$ factor; for the
second stage it is connected and no further topological defects
are formed. (Monopoles are also formed in the first stage of
symmetry breaking.) Our motivation for studying this problem is
three fold. Firstly, the non-abelian domain walls produced in the
first symmetry breaking have many unusual properties compared to
standard $Z_2$ domain walls.  As discussed in
Refs.~\cite{Pogosian:2000xv,TVorig,PV,TVsymm}, in fact three
distinct classes of domain walls are formed in this phase
transition (see also Refs.~\cite{PVlat,AntunesPV,AntunesV} for a
discussion of the resulting lattice of domain walls). Only one of
these walls is stable and the full $SU(5)\times Z_2$ symmetry is
{\it not} restored in its core. On the contrary, the symmetry
inside the stable wall is {\it smaller} than the unbroken $SU(3)
\times SU(2) \times U(1)$ symmetry.  A second motivation is to
explore grand unified phase transitions. Our analysis shows that
many complex and diverse outcomes are likely, and the eventual
cosmological consequences of a grand unified phase transition
depend on the structure and interactions of both stable and
unstable topological defects in the model.

A final motivation is to understand how fermions and monopoles
interact with such a wall.  Here we work in the context of $SU(5)$
grand unified theory (GUT), even though this model is ruled out for
a number of reasons including proton decay.  However, other
(supersymmetric) GUTs often break down to the standard model
through $SU(5)\times Z_2$ and hence phenomena similar to those
discussed here may also occur in more realistic GUTs. In this
context, $\Hi$ breaks the first symmetry in
Eq.~(\ref{basic.symbreak}) at scales
$$
v_{\rm GUT} \simeq 10^{16} \, {\rm GeV}
$$
whereas fermions get their mass from Yukawa interactions with the
electroweak Higgs $\vphi$ at scales
$$
v_{\rm EW} \simeq 100 \, {\rm GeV}.
$$
Given this hierarchy one can neglect the backreaction of the
fermions and $\vphi$ on the GUT domain walls. However, due to the
standard couplings between $\Hi$ and $\vphi$ (see, for instance
\cite{BL,CL}), after electroweak symmetry breaking, GUT domain
walls will lead to regions in space in which the electroweak Higgs
take different values. As mentioned above, the result is that as
fermions go from one region to another they can scatter in
non-trivial ways.

Before studying this problem, it will be useful to recall the main
features of fermion scattering off the familiar $Z_2$ domain wall,
and at the same time introduce some relevant notation. The action
for this coupled system is $S(\Psi,\phi)= S_1(\phi) +
S_2(\Psi,\phi)$ where $\phi$ is a real scalar field, $\Psi$ a
Dirac fermion, and
\begin{eqnarray}
S_1(\phi) & = & \int d^4 x \left( \frac{1}{2} (\partial_\mu
\phi)^2 - \frac{\lambda}{4}(\phi^2 - \eta^2)^2 \right)
\label{standard}
\\
 S_2(\Psi,\phi) &=& \int d^4 x \left(
 \bar{\Psi}(i \dslash  + g \phi) \Psi  \right)
 \label{stand1}
 \\
 &=&  \int d^4 x \biggl(
i (\bar{\Psi}_L \dslash \Psi_L + \bar{\Psi_L^C} \dslash \Psi_L^C)
 \nonumber \\
&+&g \phi  \left[ \Psi^{\rm T}_L C \Psi_L^C - (\Psi_L^C)^\dagger C
(\Psi_L)^* \right]
 \biggr )
 \label{stand2}.
\end{eqnarray}
In Eq.~(\ref{stand2}) we have used the standard notation for GUTs and
decomposed $\Psi$ into left- and right- handed components,
introducing the charge conjugate field
\begin{equation}
\Psi =\Psi_L + \Psi_R = \Psi_L + i \gamma^2 (\psi_L^C)^*
\end{equation}
where $\Psi_{L,R} = (1 \mp \gamma^5) \Psi/2$ and
\be
 \Psi_L^C
\equiv (\Psi^C)_L \, , \; \; \Psi^C = C \gamma^0 \Psi \, , \; \;
C=i \gamma^2 \gamma^0.
\ee
The $\gamma$ matrices satisfy $\{ \gamma^{\mu} , \gamma^{\nu} \} =
2 \eta^{\mu \nu}$, $(\gamma^\mu)^{\dagger} = \gamma^0 \gamma^\mu
\gamma^0 $, and $\gamma^5 \equiv i \gamma_0 \gamma_1 \gamma_2
\gamma_3$.  We work in the Dirac representation;
\begin{equation}
\gamma^0 =  \protect\left(
\begin{array}{cc}
1 &  0  \\
0 &  -1 \\
\end{array} \protect\right) , \
\gamma^i =  \protect\left(
\begin{array}{cc}
0 &  \sigma^i  \\
-\sigma^i &  0 \\
\end{array} \protect\right) , \
\gamma^5 =  \protect\left(
\begin{array}{cc}
0 \; \; &  1 \\
1 \; \; &  0 \\
\end{array} \protect\right).
\end{equation}
From Eqs.~(\ref{stand1}) and (\ref{stand2}), the fermion equation of
motion is
\begin{equation}
i\dslash \Psi +g \phi \Psi = 0
\label{basic.diraceq}
\end{equation}
or equivalently
\begin{equation}
i\dslash \Psi_L - g \phi \gamma^0 C (\Psi_L^C)^* =  0, \qquad i
\dslash \Psi_L^C -g \phi \gamma^0 C \Psi_L^* = 0.
\end{equation}

If backreaction is ignored, the domain wall solution centered on
$y=0$ and interpolating between the two discrete vacua at $\phi =
\pm \eta$ is given by the solution
\begin{equation}
\phi_{DW}(y) = \eta \tanh\left(\tilde{\sigma} y \right)
\label{DW}
\end{equation}
where $\tilde{\sigma} = \lambda \eta^2/\sqrt{2}$. At the center of
the wall $\phi_{DW}=0$ and the $Z_2$ symmetry is restored. There
exists a fermion zero mode that is localized on the wall
\cite{ViSh}. Similar comments will not hold for $SU(5)\times Z_2$
domain walls. Fermions may also scatter off the wall, in which
case the asymptotic states for Eq.~(\ref{basic.diraceq}) consist of
incoming and reflected positive energy (spin up, say) plane waves
at $y=-\infty$;
\begin{equation}
\Psi_{I} = e^{-i(\omega t-ky)} u_\uparrow(k,m)\; , \
\Psi_{R} = \alpha e^{-i(\omega t+ky)} u_\uparrow(-k,m)
\end{equation}
where $m=g\eta$,  and a transmitted wave at $y=+\infty$
\begin{equation}
\Psi_{T} = \beta e^{-i(\omega t-ky)} u_\uparrow(k,-m).
\end{equation}
Here $\alpha$ and $\beta$ are $c$-numbers, and $\omega^2 - k^2 =
m^2$. The normalized spinor $u_\uparrow(k,m)$ satisfies $(\kslash
- m) u_\uparrow(k,m)=0$ so that $u_\uparrow(k,m)^{\rm T} =
\sqrt{\omega +m}(1 , 0,0, ik/(\omega +m))$. By squaring
Eq.~(\ref{basic.diraceq}) and rewriting in terms of the variable
$2 z = 1 - \tanh (\tilde{\sigma} y)$, one obtains a hypergeometric
equation that can be solved exactly. The result may be found in
\cite{McLerran93,Farrar:1994vp}, and one can calculate for
example, the reflection and transmission coefficients
\begin{equation}
R = \left| \frac{ j^{(2)}_{R}}{j^{(2)}_{I}} \right| , \
T = \left| \frac{j^{(2)}_{T}}{j^{(2)}_{I} } \right|
\end{equation}
where the current $j^{(\mu)}$ is defined as the expectation value
of the normal-ordered quantum operator $: \bar{\Psi} \gamma^\mu
\Psi :$. In the limit of a zero thickness wall, Eq.~(\ref{DW})
reduces to
\begin{equation}
\phi_{DW}(y) = 2 \eta \biggl [ \Theta(y) -\frac{1}{2} \biggr ] \ ,
\end{equation}
and $R$ and $T$ are straightforward to calculate. The solution of
the Dirac equation, Eq.~(\ref{basic.diraceq}), is obtained by matching
the plane waves across the wall,
$u_\uparrow(k,m) + \alpha u_\uparrow(-k,m)= \beta u_\uparrow(k,-m)$,
leading to
\begin{equation}
R=\frac{m^2}{\omega^2} \, , \qquad T =
\frac{\omega^2-m^2}{\omega^2}.
\end{equation}

In the remainder of this paper we tackle a similar problem for
$SU(5) \times Z_2$ domain walls. In Sec.~\ref{dwinsu5z2} we
discuss the stable domain wall solutions in the $SU(5)\times Z_2$
model, formed in the first stage of symmetry breaking
(Eq.~(\ref{basic.symbreak})). In Sec.~\ref{eweffect} we describe
the effects of electroweak symmetry breaking on the domain wall,
and how this can lead, for example, to chromomagnetic vortices on
the wall. Fermion scattering is discussed in
Sec.~\ref{fermionscattering}.

\section{Domain walls in $SU(5)\times Z_2$}
\label{dwinsu5z2}

In this section we briefly recall the main results of
\cite{Pogosian:2000xv,TVorig,PV} regarding the properties of
domain walls formed in the first symmetry breaking
Eq.~(\ref{basic.symbreak}). Electroweak symmetry breaking will be
considered in the following section. The Langrangian density
is\footnote{Our notation is the following. The $a=1,\ldots,24$
hermitian and traceless generators of $SU(5)$ are denoted by
$\T_a$, and they are normalized such that $\Tr(\T_a \T_b) =
\delta_{ab}/2$.  We often write $\Hi = H^a \T_a$. A group element
of $SU(5)$ is denoted by $\g = \exp(i \epsilon_a \T_a)$ with
det$\g=1$ and $\g \g^\dagger ={\bf 1}$.}
\begin{equation}
{\cal{L}}_\Hi = \Tr [ (\D_\mu \Hi) (\D^\mu \Hi)^\dagger ] - V(\Hi)
+ \frac{1}{2} \Tr [F_{\mu \nu} F^{\mu \nu}]
\label{ActionH}
\end{equation}
where $\D_\mu \Hi = \partial_\mu \Hi + i e [\A_\mu,\Hi]$ and
$F_{\mu \nu} =  \partial_\mu \A_{\nu} - \partial_\nu \A_{\mu} + ie
[\A_\mu,\A_\nu]$. The potential $V(\Hi)$ is taken to be quartic in
the adjoint Higgs $\Hi$;
\begin{equation}
V(\Hi) = -m_1^2 {\rm tr} (\Hi^2) + \lambda_1 ({\rm tr} \Hi^2)^2 +
\lambda_2 {\rm tr} (\Hi^4) + V_0 \label{potV}
\end{equation}
with $V_0$ chosen such that at its minimum, $V=0$. Action
(\ref{ActionH}) is invariant under local $SU(5)$ transformations
\begin{eqnarray}
\Hi &\stackrel{SU(5)}{\longrightarrow} & \Hi'= \g\Hi \g^\dagger
\label{su5}
\\
(\D_\mu \Hi) &\stackrel{SU(5)}{\longrightarrow} & (\D_\mu \Hi)'=
\g (\D_\mu \Hi) \g^\dagger \label{Deq}
\\
\A_{\mu} &\stackrel{SU(5)}{\longrightarrow}& \A'_{\mu} = \g
\A_{\mu} \g^{\dagger} - \frac{i}{e} (\partial_\mu \g) \g^{\dagger}
\label{Aeq}
\end{eqnarray}
as well as $Z_2$ transformations
\begin{equation}
\Hi \stackrel{Z_2}{\longrightarrow} \Hi''= - \Hi \label{z2}
\end{equation}
which leave the gauge sector invariant. Notice that this $Z_2$
transformation is {\it not} included in the $SU(5)$
transformation: for any $g \in SU(5)$ one has $\Tr({\Hi'}^3) =
\Tr(\g\Hi^3 \g^\dagger) = \Tr(\Hi^3)$, whilst under $Z_2$
$\Tr (\Hi^3 )$ changes sign. Thus the action
Eq.~(\ref{ActionH}) is $SU(5)\times Z_2$ invariant provided there
are no cubic terms in the potential.\footnote{One should note a
subtlety regarding this $Z_2$ transformation.  There are certain
field configurations, $\Hi_s$, satisfying ${\rm Tr}(\Hi_s^{2n+1})=0$
with $n \in {\cal Z}$ for which one {\it can} change the sign of $\Hi_s$
using an $SU(5)$ transformation.  An example is $\bar{\Hi}_s
\equiv {\rm diag}(1,-1,0,0,0)$: for $\g = (\sigma_1,1,1,-1)$,
$\bar{\Hi}_s' = \g \bar{\Hi}_s \g^\dagger=-\bar{\Hi}_s$.}

As discussed in e.g.~\cite{BL,CL}, the choice of parameters
\begin{equation}
\lambda_2 \geq 0, \qquad \frac{\lambda_1}{\lambda_2} \geq
-\frac{7}{30}, \qquad m_1^2 > 0
\end{equation}
breaks $SU(5) \times Z_2 \longrightarrow SU(3)\times SU(2) \times
U(1)/Z_6$. Following \cite{PV}, the domain wall solution is known
analytically if
\begin{equation}
\lambda_1 = -3 \lambda_2/20.
 \label{l1l2}
\end{equation}
The potential is minimized for
\begin{equation}
\Hi_0 = v_{GUT} \; {\rm diag}(2,2,2,-3,-3) \label{vevH}
\end{equation}
where
\begin{equation}
v_{GUT}^2 =\frac{ m_1^2}{5\lambda_2},
\end{equation}
and $\Hi_0$ is invariant under the unbroken subgroup
\begin{equation}
K_0 = SU(3)\times SU(2) \times U(1)/Z_6
\end{equation}
so that
\begin{equation}
\Hi_0 = {\bf h} \Hi_0 \h^{\dagger} \qquad \forall \; { {\bf h} }
\in K_0.
\end{equation}
It is straightforward to identify the (8+3+1)=12 generators of
$K_0$ out of the 24 generators of $SU(5)$ (see e.g.~\cite{BL}).

\begin{figure}
\centerline{\scalebox{1.00}{\input{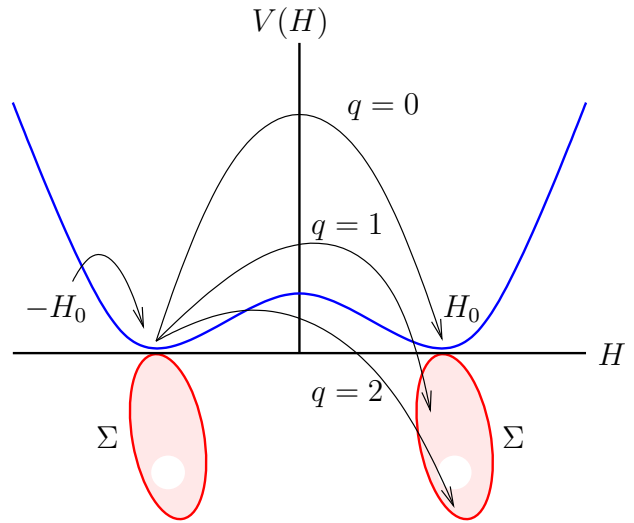}}}
\caption{The vacuum manifold for $SU(5)\times Z_2$ breaking is in
two identical disconnected pieces, each labeled by $\Sigma$.
$\Sigma$ itself has non-trivial topology, including holes which
lead to $\Pi_2(\Sigma) \neq 0$ and hence monopoles. Different
domain wall solutions, labeled by $q=0,1,2$, are obtained for
different boundary conditions at spatial infinity. }
\label{vacmnfld}
\end{figure}

The vacuum manifold of the potential given in Eq.~(\ref{potV}) has
two discrete sectors linked by the $Z_2$ transformation.
Furthermore, each sector consists of a whole manifold of vacua
generated by $SU(5)$ transformations (see Fig.~\ref{vacmnfld} for
a schematic representation). This should be contrasted with the
standard potential of Eq.~(\ref{standard}).
Correspondingly, the properties of the resulting domain walls are
more diverse.

By symmetry, the only non-trivial component of the gauge field in
the static domain wall will be $\A^y$. This can always be set to
zero by a gauge transformation and hence in the following
discussion we suppose that all gauge fields vanish.

Suppose one fixes
\begin{equation}
\Hi(y\rightarrow -\infty) = \Hi_0.
\end{equation}
Then topologically non-trivial boundary conditions occur for
$\Hi(y \rightarrow \infty) \equiv \Hi^+$ given by
\begin{equation}
\Hi^+ = - \g\Hi_0 \g^\dagger \qquad \forall \; \g \in SU(5).
\label{top}
\end{equation}
Suppose $\g \in K_0$. Then $\Hi^+ = -\Hi_0$ (see
Fig.~\ref{vacmnfld}), but the resulting ``$q=0$'' domain wall is
unstable \cite{Pogosian:2000xv,TVorig,PV}.  The only stable domain
wall lies in the ``$q=2$'' class \cite{Pogosian:2000xv,TVorig,PV}
for which the boundary condition at $+\infty$ is
\begin{equation}
\Hi_{(q=2)}^+ = - v_{\rm GUT} \; \; {\rm diag}(2,-3,-3,2,2).
\label{Hi+q=2}
\end{equation}
The explicit domain wall solution can be written in closed form
for the choice of parameters in Eq.~(\ref{l1l2}):
\begin{eqnarray}
\Hi_{(q=2)}(y) &=& \frac{v_{GUT}}{2} \biggl[ \tanh (\sigma y)
~ {\rm diag} (-4,1,1,1,1) \nonumber \\
&&\qquad + ~ {\rm diag}(0,5,5,-5,-5) \biggr]
\label{Hq2}
\\
\A^{\mu}_{(q=2)}(y) &=& 0.
\end{eqnarray}
Observe that the full $SU(5)$ symmetry is {\it not} restored at
the center of the wall since the unbroken symmetry there is
$SU(2)^2 \times U(1)^2$, which is smaller than the unbroken
$SU(3)\times SU(2) \times U(1)$.  In the wall color symmetry is
broken: $SU(3) \rightarrow SU(2) \times U(1)/Z_3$.

We now focus on these ``$q=2$'' walls and determine their effect on
the electroweak Higgs $\vphi$ and hence on electroweak symmetry
breaking.

\section{Effect of electroweak symmetry breaking on the domain wall}
\label{eweffect}

So far we have considered the GUT symmetry breaking which is the
first part of Eq.~(\ref{basic.symbreak}). Now we turn to the
second part of Eq.~(\ref{basic.symbreak}) in which the electroweak
symmetry breaks due to the VEV of the scalar field, $\vphi$.
In the next two subsections, we will first discuss the bosonic
sector, focusing on the solution for $\vphi$ in the background of
the ``$q=2$'' GUT domain wall. Then we discuss the fermionic sector in
the background of the domain wall which now includes the solution
for $\vphi$. Scattering of fermions off the domain wall is
considered section \ref{fermionscattering}.

\subsection{Bosonic sector}

The Lagrangian density for the bosonic sector of the GUT theory is
\cite{BL}
\begin{equation}
{\cal L} = {\cal L}_{\Hi} + (\D_\mu \vphi)(\D^\mu \vphi)^\dagger
-V(\vphi) - V(\Hi,\phi)
\label{Actionphi}
\end{equation}
where ${\cal L}_{\Hi} $ is given in Eq.~(\ref{ActionH}), and
\begin{eqnarray}
V(\vphi) &=& -m_2^2(\vphi^\dagger \vphi) + \lambda_3
(\vphi^\dagger \vphi)^2
\label{Vp}
\\
V(\Hi,\vphi) &=& \lambda_4 ({\rm tr} \Hi^2)(\vphi^\dagger \vphi) +
\lambda_5 (\vphi^\dagger \Hi^2 \vphi) \label{Vhp}
\end{eqnarray}
with $m_2^2>0$ and $\lambda_3>0$. Under $SU(5)$ and $Z_2$
\begin{equation}
\vphi \stackrel{SU(5)}{\longrightarrow}  \vphi' = \g \vphi \; ,
\qquad \vphi \stackrel{Z_2}{\longrightarrow}  \vphi'' = +\vphi.
\end{equation}
Note that the Lagrangian is also invariant under the transformation
$(\Hi, \vphi) \rightarrow (\Hi,-\vphi)$ for any $(\Hi, \vphi)$.
This is an additional $Z_2'$ symmetry of the model. However,
the $Z_2'$ symmetry, in combination with $SU(5)$ rotations,
remains unbroken throughout, and we shall ignore it in the
remainder of this paper.

Given the domain wall solution of Eq.~(\ref{Hq2}), one can
determine $\vphi$ in this background. From Eqs.~(\ref{Vp}) and
(\ref{Vhp}), the effective potential for $\vphi$ is
\begin{eqnarray}
\tilde{V}(|\vphi|) &=& V(\vphi) + V(\Hi,\vphi) \nn
\\
&=& |\phi_1|^2 \mu_1^2(y) + \left(|\phi_2|^2 + |\phi_3|^2 \right)
\mu_{23}^2(y) \nonumber \\
&+& \left(|\phi_4|^2 + |\phi_5|^2 \right) \mu_{45}^2(y)
+ \lambda_3  \sum_{p=1}^5 \sum_{q \neq p} |\phi_p|^2 |\phi_q|^2
\nonumber \\
&+& \lambda_3 \sum_{p=1}^5 |\phi_p|^4 \label{potphi}
\end{eqnarray}
where $p=1,\ldots,5$ labels the components of $\vphi$ and
\begin{eqnarray}
\mu_1^2(y) &=&
-m_2^2 + v_{\rm GUT}^2
\left[ 25 \lambda_4 + \left( 5 \lambda_4 + 4 \lambda_5 \right)
\tanh^2(\sigma y) \right] \; \; \; \; \label{mu1}
\\
\mu_{23}^2(y) &=& -m_2^2 + v_{\rm GUT}^2 \biggl[ 25 \left(\lambda_4
+ \frac{\lambda_5}{4} \right)  \nonumber \\
&+& 5 \left( \lambda_4 + \frac{\lambda_5}{20} \right)\tanh^2(\sigma y)
+ \frac{5 \lambda_5}{2} \tanh(\sigma y) \biggr ] \label{mu23}
\\
\mu_{45}^2(y) &=& \mu_{23}^2(-y) \label{mu45}.
\end{eqnarray}
Asymptotically
\begin{eqnarray}
\mu_1^2(\pm \infty) &=& \mu_{23}^2(-\infty) = \mu_{45}^2(+\infty)
\nonumber \\
&=& -m_2^2 + v_{\rm GUT}^2 \left[ 30 \lambda_4 + 4 \lambda_5 \right]
\equiv \mu^2_{SU(3)} \label{mu3}
\\
\mu_{23}^2(+\infty) &=& \mu_{45}^2(-\infty) \nonumber \\
&=& -m_2^2 + v_{\rm GUT}^2
 \left[ 30 \lambda_4 + 9
\lambda_5 \right] \equiv \mu^2_{SU(2)}. \label{mu2}
\end{eqnarray}
and from Eqs.~(\ref{vevH}) and (\ref{Hi+q=2}) the boundary conditions
are
\begin{equation}
\vphi(y \rightarrow -\infty) =  \protect\left( \begin{array}{c}
0  \\
0  \\
0  \\
0  \\
v_{\rm EW}  \\
\end{array} \protect\right),
\qquad \vphi (y \rightarrow + \infty)=  \protect\left(
\begin{array}{c}
0  \\
0  \\
v_{\rm EW}  \\
0  \\
0  \\
\end{array} \protect\right)
\label{bconphi}
\end{equation}
with associated unbroken U(1) generators
\begin{eqnarray}
Q(y \rightarrow -\infty) &=& {\rm diag}(-1/3,-1/3,-1/3,1,0)
\label{Qminus}
\\
 Q(y \rightarrow +\infty) &=& {\rm diag}(-1/3,1,0,-1/3,-1/3)
 \label{Qplus}
\end{eqnarray}

These boundary conditions are obtained from the potential in
Eq.~(\ref{potphi}) provided
\begin{equation}
\mu^2_{SU(3)} + 2  \lambda_3 v_{\rm EW}^2 > 0 \; , \qquad
\mu^2_{SU(2)}< 0 \label{su3mass}
\end{equation}
so that
\begin{equation}
v_{\rm EW}^2 = \frac{|\mu^2_{SU(2)}|}{2\lambda_3} =-
\frac{\mu^2_{SU(2)}}{2\lambda_3}.
\end{equation}
Furthermore, from Eq.~(\ref{su3mass}) $\mu^2_{SU(3)} -
\mu^2_{SU(2)} > 0$ and hence Eqs.(\ref{mu3}) and (\ref{mu2}) imply
that $\lambda_5 < 0.$
If one assumes that there are no fine tunings between the coupling
constants, the hierarchy
$v_{\rm EW}^2/ v_{\rm GUT}^2 \equiv \epsilon \simeq 10^{-24}$
implies that
\cite{BL,CL},
\begin{equation}
\lambda_a \equiv \lambda_4 + \frac{3}{10} \lambda_5 > 0
\label{lambdaa}
\end{equation}
so that $\lambda_4>0$.  Finally, it will be useful to define
\begin{equation}
r= - \frac{\lambda_5}{\lambda_a}.
\end{equation}
In a realistic scenario the different coupling constants are
temperature dependent and the constraints apply to the effective
coupling constants.

\subsubsection{$\phi_1$ condensate and monopole interactions}

\begin{figure}
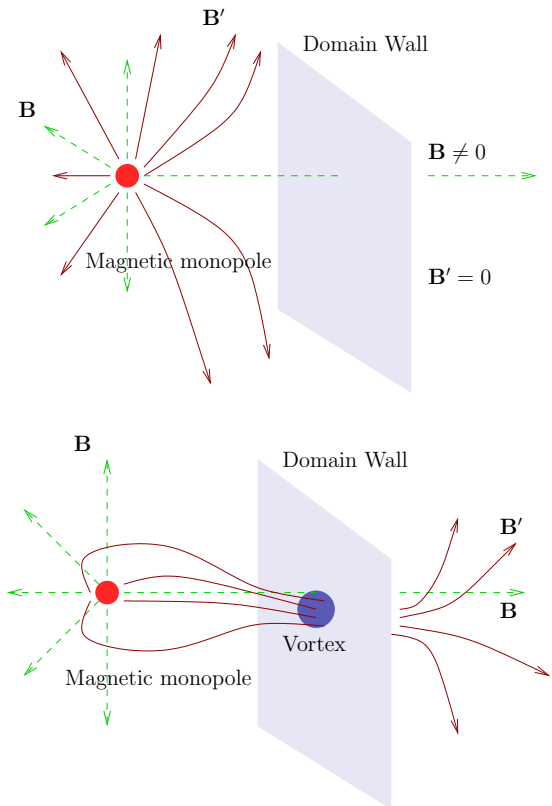

\centerline{\scalebox{0.70}{\input{floating_monopole.pstex_t}}}
\vskip 0.5cm
\centerline{\scalebox{0.70}{\input{stuck_monopole.pstex_t}}}
\caption{On the wall a combination of the $SU(3)$ and hypercharge
magnetic fields (denoted by ${\bf B}'$) is massive due to the
$\phi_1$ condensate. A magnetic monopole sourcing ${\bf B}'$
is repelled from the wall due to the Meissner effect, as shown in
the top figure. Alternatively, the ${\bf B}'$ magnetic flux passes
through the wall in a flux tube and the monopole is attracted
to the domain wall, eventually being pulled
through the wall (second figure). The orthogonal components of the
monopole magnetic field, denoted by ${\bf B}$, do not interact
with the wall.}
\label{monopolewall}
\end{figure}

Given these signs of coupling constants, there may be a condensate
of $\phi_1$ on the wall (both before and after the electroweak
phase transition). Indeed, from Eq.~(\ref{mu1}), $\mu_1^2(0) < 0$
provided $r<2/5$.  This condensate transforms non-trivially under
$SU(3)$ color and $U(1)$ hypercharge symmetries, and hence leads
to non-trivial interactions with any magnetic monopoles also
carrying the same charges.  Let us denote the relevant magnetic
flux by ${\bf B}'$ (see Fig.~\ref{monopolewall}).  If the
$\phi_1$ condensate is uniform, the monopole can be repelled from
the wall through the Meissner effect.  The force can be found
using the method of images and is inversely proportional to the
square of the distance between the monopole and the wall. However,
there is also another possibility: since $\phi_1$ is a complex
field, and since the condensate is localized on the wall, there
could be vortices situated on the wall due to the winding of
$\phi_1$. (This is the analogue of a ``boojum'' in condensed
matter physics
--- that is, a defect that can only live on the surface of a
container because of the boundary conditions.) Now, depending on
the relative orientation of the condensate and the magnetic
monopole charge, the magnetic field due to the monopole could pass
through the wall in quantized units of flux in the form of a
vortex.

Another intriguing point linked to this condensate is that it
implies, at least classically, the existence of massive fermions
on the wall, since the fermions get their mass from interactions
with the electroweak Higgs (see Sec.~\ref{sec:fermions}).  Thus,
before the electroweak phase transition it would seem to be
possible to have massless fermions off the wall but massive
fermions confined on it.

We hope to return to a more detailed investigation of both these
possibilities in the future.

\subsubsection{No $\phi_1$ condensate}

In the remainder of this paper we focus on the case in which
$r>2/5$ so that $\phi_1(y)=0$.  The fields $\phi_2$ and $\phi_4$
can be set to zero so that the remaining degrees of freedom are
$\phi_{3,5}$, and thus the profile of $\vphi$ is
\begin{equation}
\vphi (y)=  \protect\left(
\begin{array}{c}
0  \\
0  \\
\alpha(y) \\
0  \\
\beta(y)  \\
\end{array} \protect\right) \equiv \sqrt{\alpha^2 + \beta^2} \protect\left(
\begin{array}{c}
0  \\
0  \\
\cos\theta \\
0  \\
\sin\theta  \\
\end{array} \protect\right)
\end{equation}
where the asymptotic behavior of $\alpha(y)$ and $\beta(y)$ is
fixed by Eq.(\ref{bconphi}), and
\begin{equation}
\tan \theta  =
\frac{\beta (y)}{\alpha (y)}
\end{equation}
so that
\begin{equation}
\theta(-\infty)=\frac{\pi}{2} \; , \qquad \theta(+\infty)=0.
\label{bctheta}
\end{equation}

For the following analysis it will be convenient to change gauge
and work in a gauge in which $\vphi(y)$ is non-zero only in the
5th entry. Consider therefore $\g(y) \in SU(5)$ given by
\begin{equation}
\g(y) = \protect\left(
\begin{array}{ccccc}
1 \; \; \; \;&                    0 &                    0 &                      0 & 0 \\
0 \; \; \; \;&\sin \theta  &                    0 & -\cos \theta  & 0 \\
0 \; \; \; \;&                    0 & \sin \theta & 0                       & -\cos \theta \\
0 \; \; \; \;& \cos \theta &                    0 &\sin \theta     & 0 \\
 0 \; \; \; \;& 0                    & \cos \theta &                      0 & \sin \theta \\
\end{array} \protect\right)
\label{thatsit}
\end{equation}
 with
$\g(-\infty) = {\rm diag}(1,1,1,1,1)$. Then after a gauge
transformation ($\vphi(y) \rightarrow \g(y) \vphi(y)$),
\begin{equation}
\vphi(y) = \sqrt{\alpha(y)^2+\beta(y)^2}  \protect\left(
\begin{array}{c}
0  \\
0  \\
0 \\
0  \\
1  \\
\end{array} \protect\right),
\label{phinew}
\end{equation}
and
\begin{equation}
\A_{y} = \frac{i}{e}\frac{d\theta}{dy} \protect\left(
\begin{array}{ccccc}
0 \; \; & 0 & 0 & \; \; \; 0 & \; \; \; \;0 \\
0 \; \; & 0 & 0& \; \; \; 1 & \; \; \; \;0 \\
0 \; \; & 0 & 0 & \; \; \; 0 & \; \; \; \;1 \\
0 \; \; & -1 & 0 & \; \; \; 0 & \; \; \; \;0 \\
0 \; \; & 0 & -1 & \; \; \; 0 & \; \; \; \;0 \\
\end{array} \protect\right).
\label{vevp}
\end{equation}
The gauge transformation (Eq.~(\ref{su5})) of $\Hi$ yields
\begin{equation}
\Hi(-\infty) = \Hi_0 =- \Hi(+\infty) \label{Hpmm}
\end{equation}
and finally, in this gauge, $Q(-\infty) = Q(+\infty)$ as given by
Eq.~(\ref{Qminus}). Note that in the thin wall approximation,
which we shall be using later,
\begin{equation}
\alpha(y) =v_{\rm EW} \Theta(y), \; \beta(y) =v_{\rm EW}
\Theta(-y),
\end{equation}
 and
\begin{equation}
\frac{d\theta}{dy} \simeq -\frac{\pi \delta(y)}{2}.
\label{gaugedeltafn}
\end{equation}
Furthermore, in this limit where $\Hi (y) \propto \Hi_0$, one can
see from Eq.~(\ref{Qminus}) that $\A_y$ carries electromagnetic
charge since $[ Q , \A_y ] \ne 0$. Similarly, it can also carry
color charge (indeed $\A_y$ is a lepto-quark boson \cite{CL}).
Finally, we note that the $\g(y)$ of Eq.~(\ref{thatsit}) is not
unique: there exist many other gauge transformations for which
Eqs.~(\ref{phinew}) and (\ref{Hpmm}) are also satisfied. Some of
these transformations may have the advantage of yielding gauge
fields with definite charges.  Despite that we use the simpler
transformation Eq.~(\ref{thatsit}), and now turn to the fermion
equations of motion.

\subsection{Fermions}
\label{sec:fermions}

The Lagrangian for the fermionic sector is \cite{BL,CL}
\begin{eqnarray}
{{\cal L}}_f &=& i (\bar{\psi}_p)_L \Dslash ({\psi}_p)_L + i
\bar{\chi}^{pq}_L \Dslash {\chi}^{pq}_L
\nn
\\ &+& g_1 \left[ (\psi_p)_L^{\rm T} C \chi_L^{pq} \phi^\dagger_q
-  \phi_q  \bar{\chi}_L^{pq} \gamma^0 C (\psi_p)_L^* \right]
\label{mm1}
\\ &+& g_2 \epsilon_{pqrst} \left[  (\chi_L^{pq})^{\rm T} C
\chi^{rs}_L \phi^t - (\phi^t)^\dagger
\bar{\chi}_L^{pq} \gamma^0 C (\chi^{rs}_L)^* \right]
\nonumber
\end{eqnarray}
where $\psi$ is a vector transforming as
$\bar{5}$, and $\chi$ is an antisymmetric tensor $10$, so that the
Lagrangian is invariant under $\vpsi \rightarrow \g^* \vpsi$,
$\chi \rightarrow  \g \chi \g^{\rm T}$ and $\vphi \rightarrow  \g
\vphi$. The equations of motion are
\begin{equation}
i [\dslash ({\psi}_p)_L - i e \gamma^2 (\A_2^*)_{pq}
({\psi}_q)_L ] - g_1 \gamma^0 C (\chi_L^{pq})^* \phi_q = 0
\label{ff1b}
\end{equation}
\begin{eqnarray}
i[\dslash {\chi}^{pq}_L + 2 i e \gamma^2 (\A_2)_{pr}
{\chi}^{rq}_L ]
+ \frac{ g_1}{2}  \gamma^0 C \left[ \phi_p (\psi_q)_L^*
- \phi_q (\psi_p)_L^* \right]
&&
\nonumber \\
-2 g_2 \epsilon_{rspqt} (\phi^t)^\dagger \gamma^0 C
                                (\chi^{rs}_L)^* = 0 \hskip 1cm &&
\label{ff2b}
\end{eqnarray}
The 15 degrees of freedom are accommodated as follows
\begin{equation}
(\psi_p)_L =  \protect\left(
\begin{array}{c}
d_1^C   \\
d_2^C  \\
d_3^C  \\
e  \\
\nu  \\
\end{array} \protect\right)_L
\end{equation}
\begin{equation}
\chi_L^{pq} =
\frac{1}{\sqrt{2}}\protect\left(
\begin{array}{ccccc}
0& u_3^C & -u_2^C & -u_1 & -d_1  \\
-u_3^C & 0 & u_1^C & -u_2 & -d_2  \\
u_2^C & -u_1^C & 0 & -u_3 & -d_3  \\
u_1& u_2 & u_3 & 0 & -e^C  \\
d_1& d_2 & d_3 & e^C & 0  \\
\end{array} \protect\right)_L
\end{equation}
and the electron and quark masses are
\begin{equation}
m_e = m_d = \frac{g_1 v_{{\rm EW}}}{\sqrt{2}} \qquad
m_u = 4 g_2 v_{{\rm EW}}.
\end{equation}

\section{Scattering fermions off the wall}
\label{fermionscattering}

We now use the equations of motion Eq.~(\ref{ff1b}) and
Eq.~(\ref{ff2b}) to determine how fermions scatter off the $q=2$
domain wall.  It is straightforward to see that the equations for
$(\psi_{5,3})_L$ and $(\chi_{35})_L$ (that is, $\nu_L$ and $d_3$)
are coupled, but decoupled from the other fermionic components.
Hence initially we focus on these. Eq.~(\ref{ff2b}) with
$(p,q)=(3,5)$ yields
\begin{equation}
i \dslash \chi^{35}_L - \frac{m_d}{\sqrt{2}} \gamma^0 C
(\psi_3)^*_L   = 0 \label{d3delta}
\end{equation}
while Eq.~(\ref{ff1b}) with $p=5$ and $p=3$ gives, respectively
\begin{equation}
i \dslash (\psi_5)_L + e \gamma^2 (\A_2^*)_{53} ({\psi}_3)_L  = 0,
\label{dd}
\end{equation}
\begin{equation}
i\dslash ({\psi}_3)_L + e \gamma^2 (\A_2^*)_{35} ({\psi}_5)_L  -
\sqrt{2} m_d \gamma^0 C (\chi_L^{35})^* = 0. \label{nudelta}
\end{equation}

In the limit of a zero thickness wall, the gauge field is a
$\delta-$function centered on the wall (see
Eq.~(\ref{gaugedeltafn})), and Eqs.~(\ref{dd}) and (\ref{nudelta})
give the matching conditions across the wall. Using (\ref{vevp}),
we obtain
\begin{eqnarray}
\frac{d(\psi_5)_L}{d\theta} + (\psi_3)_L &=& 0 \nn
\\
\frac{d(\psi_3)_L}{d\theta} - (\psi_5)_L &=& 0, \nn
\end{eqnarray}
so that on integrating and imposing the boundary conditions
Eq.~(\ref{bctheta}) we find
\begin{eqnarray}
\left. (\psi_3)_L \right|_{0^-} &=& \left. (\psi_5)_L \right|_{0^+} ,
\label{bc1}
\\
\left. (\psi_5)_L \right|_{0^-} &=& -\left. (\psi_3)_L
\right|_{0^+}. \label{bc2}
\end{eqnarray}
From Eq.~(\ref{d3delta}), $\chi^{35}_L$ is continuous across the
wall. Suppose a neutrino is incident on the wall from $-\infty$,
$\left. (\psi_5)_L \right|_{0^-} \neq 0$. Then from
Eq.~(\ref{bc2}), $\left. (\psi_3)_L \right|_{0^+} \neq 0$ so that
one expects a transmitted down quark.  It is straightforward to
solve this scattering problem.  In the Dirac representation and
using the projector $(1-\gamma^5)/2$, the incident left-handed
neutrino is given by (see Eq.~(\ref{dd}))
\be
(\psi_5)^i_L(y<0) = e^{-i\omega (t-y)} \protect\left(
\begin{array}{c}
1  \\
-i  \\
-1 \\
i \\
\end{array} \protect\right).
\ee
while the reflected neutrino and the different left handed
components of the down quark are given by
\ba
(\psi_5)^r_L(y<0) &=& \alpha_r e^{-i\omega (t+y)} \protect\left(
\begin{array}{c}
1  \\
i  \\
-1 \\
-i \\
\end{array} \protect\right)
\\
(\psi_3)_L^{r}(y<0) &=&  e^{-i(\omega t+k'y)} \protect\left(
\begin{array}{c}
p  \\
q  \\
-p \\
-q \\
\end{array} \protect\right),
\\
(\chi^{35}_L)^{r*}(y<0) &=&  \frac{1}{\sqrt{2}}
          e^{-i(\omega t+k'y)} \protect\left(
\begin{array}{c}
r  \\
s  \\
-r \\
-s \\
\end{array} \protect\right)
\ea
where, from Eqs.~(\ref{d3delta}) and (\ref{nudelta}), $\omega^2 -
k'^2 = m_d^2$ and
\be
s = \frac{1}{m_d}(\omega p +ik'q) \; , \; r = -
\frac{1}{m_d}(\omega q-ik'p).
\ee
For $y>0$, the transmitted waves are
\ba
(\psi_5^L)_{t}(y>0) &=& \alpha_t e^{-i\omega(t-y)} \protect\left(
\begin{array}{c}
1  \\
-i  \\
-1 \\
i \\
\end{array} \protect\right)
\\
(\psi_3^L)_{t}(y>0) &=&  e^{-i(\omega t-k'y)} \protect\left(
\begin{array}{c}
f  \\
g  \\
-f \\
-g \\
\end{array} \protect\right)
\\
(\chi^{35}_L)^{t*}(y>0) &=&  \frac{1}{\sqrt{2}}
e^{-i(\omega t-k'y)} \protect\left(
\begin{array}{c}
a  \\
b \\
-a \\
-b \\
\end{array} \protect\right)
\ea
where
\begin{equation}
 b = \frac{1}{m_d}(\omega f-ik'g) \; , \; a=-  \frac{1}{m_d}(\omega
 g+ik'f).
\end{equation}
From the boundary conditions Eqs.~(\ref{bc1}) and (\ref{bc2}) we find
\begin{equation}
\alpha_r = 0 \ , \; \; \alpha_t = {\frac { \left( \omega -k'
\right) }{\omega +k'}}
         = p = iq
\end{equation}
\begin{equation}
f=-1=i g
\end{equation}
\begin{equation}
r = - i{\frac {\left( \omega -k' \right) }{m_d}}  = is = a = ib.
\end{equation}

Since $\alpha_r=0$ there is no reflected neutrino. This could
have been expected as a consequence of helicity conservation and
the fact that the interaction of the domain wall does not depend
on the spin orientation of the fermion.
The incoming neutrino scatters either by reflection as a
down quark, or gets transmitted as a neutrino or a down
quark. Such lepton (and baryon) non-conserving processes
have been studied in a number of other situations such as
fermion scattering off $SU(5)$ monopoles
\cite{Dokos:1979vu,Rubakov:1982fp,Callan:1982au}. Furthermore,
notice that {\it all} the transmitted and reflected waves are
proportional to the same spinor $\xi \equiv (1,-i,-1,i)^{\rm T}$,
which is a left-handed eigenstate of the {\it massless} Dirac
equation.  The (massive) down quark spinors are also
proportional to $\xi$, reflecting the fact that they are not
momentum eigenstates --- translational symmetry is
broken by the presence of the wall.

This scattering process does not conserve electric charge: the
incident electric current vanishes, while the transmitted and
reflected electric currents are {\it not} equal and opposite.  We
find
\begin{equation}
j_{r} = - \frac{(\omega-k')}{(\omega+k')} j_t.
\end{equation}
The scattering must therefore be accompanied by the transfer of
charge to the domain wall, and in a similar manner by a transfer
of color charge.  These different charges will be carried by the
massless gauge excitations $\A_y$ living on the wall.

Here we have focused on an incoming $\nu_L$ scattering into $d_3$.
These results can straightforwardly be generalized to both
incoming neutrinos and $d_3$'s, as well as the interactions
between other particles (such as $d_1$ and $u_2^C$ see
Eq.~(\ref{ff2b})).

\section{Conclusions}

In this paper we have studied a number of different aspects
related to the domain walls formed in the spontaneous symmetry
breaking of $SU(5)\times Z_2 \rightarrow SU(3)\times SU(2)\times
U(1)/Z_6$, and in particular the effect of the subsequent symmetry
breaking
down to $SU(3)\times U(1)/Z_3$.  As we have discussed, even before
the second symmetry breaking, the presence of a second scalar
field (the vector Higgs $\vphi$) can lead to the existence of a
scalar condensate on the wall ($\phi_1$) which is charged under
$SU(3)$ and $U(1)$. The interaction of this condensate with
magnetic monopoles also carrying $SU(3)$ and $U(1)$ charges can be
diverse. For example, a monopole can either be repelled from the
wall through a (generalized) Meissner effect; or it can be
attracted to the wall with its flux threading the wall through a
vortex configuration of $\phi_1$ on the wall.  We aim to study
these phenomena, and the seeming existence of massive fermions
localized on the wall, in more detail in the future.

After the second symmetry breaking phase transition (the
electroweak phase transition) when $\langle \vphi \rangle \neq 0$,
fermions off the wall become massive.  When a fermion scatters
with the wall, lepton and baryon numbers can change, and we have
shown explicitly that a neutrino can either scatter by reflection
into a down quark, or be transmitted together with a down quark.

The $SU(5)$ grand unified model we have used for motivation for
this work is not phenomenologically viable, and similarly a
network of domain walls is not cosmologically viable (though this
second issue can be resolved by adding a small ${\rm Tr}\Hi^3$
term to the model we have studied: the domain walls still exist
but they now eventually decay).  Despite these comments we believe
that our work is of interest as it illustrates some of the very
complex properties of grand unified models which have not been
fully probed in the past.  Furthermore the effects we have
highlighted here may appear in other models or indeed in other
systems.

\begin{acknowledgments}
We are grateful to Levon Pogosian for collaboration in the initial
stages of this project. This work was supported by the DOE and
NASA at Case, and by the University of Paris VII.
\end{acknowledgments}

\end{document}